\documentclass[a4paper,12pt]{article}
\usepackage{amsmath,amsfonts,amssymb,cite}
\newcommand{\be}{\begin{equation}}
\newcommand{\ee}{\end{equation}}

\begin{document}

\begin{center}
{\Large\bf About the possibility of five-dimensional effective theories for low-energy QCD}
\end{center}

\begin{center}
{\large S. S. Afonin\footnote{On leave of absence from V. A. Fock
Department of Theoretical Physics, St. Petersburg State
University, 1 ul. Ulyanovskaya, 198504, Russia.}}
\end{center}

\begin{center}
{\it Institute for Theoretical Physics II, Ruhr University Bochum,
150 Universit\"{a}tsstrasse, 44780 Bochum, Germany}
\end{center}

\begin{abstract}
The AdS/QCD models suggest an interesting idea that the effective
theory of low-energy QCD may be formulated as a 5-dimensional
field theory in the weak coupling regime in which the fifth
coordinate plays a role of inverse energy scale. Taking the point
of view that this is just an efficient parametrization of the
non-perturbative dynamics of strong interactions, we discuss on a
qualitative level an alternative possibility for a simpler
5-dimensional parametrization of main phenomena in the low-energy
QCD. We propose to interpret the effect of chiral symmetry
breaking as an effective appearance of compactified extra
dimension with the radius of the order of inverse scale of chiral
symmetry breaking. Following some heuristic arguments two dual
scenarios for the emergence of the excited light mesons are
introduced: In the first scenario, the meson resonances are
interpreted as the effects of Kaluza-Klein excitations of quarks
inside mesons, in the second one, as the formation of gluon
strings wound around the compactified dimension an appropriate
number of times. Matching of these scenarios permits to express
the slope of radial Regge trajectories through the order
parameters of the chiral symmetry breaking, with the
compactification radius being excluded. This example shows
qualitatively that the extra dimension may play an auxiliary role
providing a short way for deriving new relations.
\end{abstract}


\newpage

\section{Introduction}

Recently, the attempts to develop a systematic non-perturbative
approach to strongly coupled gauge theories have led to necessity
to invoke an extra spatial dimension --- it is required by the
AdS/CFT correspondence~\cite{maldacena}. In application to QCD,
the efforts to find an appropriate five-dimensional dual theory
formulated on an anti-de-Sitter (AdS) gravity background are
referred to as AdS/QCD~\cite{hw,AdS,erd}.

A genuine 5d holographic dual for QCD hardly can exist (see, {\it
e.g.}, the recent discussions in~\cite{er}), nevertheless, the
AdS/QCD approach turned out to be surprisingly successful in
description of low-energy observables. Thus, an important lesson
from studies of AdS/QCD models consists in the fact that the
low-energy dynamics of strong interactions may be parametrized
geometrically if one introduces one extra spatial dimension.

One can assume that this fact has nothing to do with the original
AdS/CFT correspondence but reflects something that we do not yet
understand. It is therefore interesting to find alternative extra
dimensional parametrizations of non-perturbative dynamics of QCD
which could in a simpler way compared to AdS/QCD describe the
hadron spectrum and related physics.

The identification of the radial excitations of light mesons with
the tower of Kaluza-Klein modes is at the heart of AdS/QCD models.
This observation can be used to simplify significantly the
5-dimensional description of meson excitations and to make it less
model dependent --- instead of warped extra dimension we will
consider a compactified extra dimension of a radius $R$, the 5d
space-time in this picture is therefore flat with the topology
$M^4\times S^1$. In addition, we will assume that interactions
in such a theory are weak to the extent that they can be neglected when
calculating the mass spectrum in the first approximation. From the
4d point of view, a free 5d particle in the space $M^4\times S^1$
looks as an infinite tower of Kaluza-Klein modes with the masses
$m_n\sim n/R$ (see, {\it e.g.}, a review~\cite{kub}), {\it i.e.}
we recover immediately the asymptotic spectrum of the hard-wall
AdS/QCD models~\cite{hw}. In reality, however, one expects the Regge
spectrum
\be
\label{001}
m^2\sim n.
\ee
In order to obtain the
relation~\eqref{001} within the AdS/QCD models one needs to assume
a rather nontrivial 5d background~\cite{AdS} (see, however,~\cite{afonin}).
The crucial step in our approach is to replace this assumption by
another one that could be justified within the real QCD. We assume
that {\it all} masses square of light mesons are linear in the
current quark mass,
\be
\label{01}
m^2\sim C_1+C_2m_q,
\ee
where
$C_1$ and $C_2$ are some constants. We will show with the help of
simple model-independent estimates that the radius $R$ is of the
order of $\Lambda_{\text{CSB}}^{-1}$, where $\Lambda_{\text{CSB}}$
is the scale of the Chiral Symmetry Breaking (CSB), about 1~GeV,
and is related to the order parameters of CSB. This leads to the
interpretation of the CSB as an effective emergence of one
compactified extra spatial dimension.

The paper is organized as follows. In Sect.~2, we provide
arguments concerning the validity of relation~\eqref{01} and
discuss the spectrum of excited states. Then we propose, in
Sect.~3 and Sect.~4, two complementary scenarios for the emergence
of linear spectrum, $m_n^2\sim n$: In the first case, the
excitations are interpreted as the Kaluza-Klein modes of quarks,
in the second one, the $n$-th excitation represents a gluon string
between the quark and antiquark that is wound $n$ times around a
compactified extra dimension. Alternative interpretations for the
compactification radius $R$ are discussed in Sect.~5. In Sect.~6,
we compare the proposed scenarios with the AdS/QCD approach and
other extra dimensional models. We conclude in Sect.~7.

\section{Meson mass vs. quark mass}

The origin of light hadrons is believed traditionally to be hidden
in the highly complicated non-perturbative strong interactions
described by QCD. The only well-established relation between the
parameters of the QCD Lagrangian and the hadron masses is the
Gell-Mann--Oakes--Renner (GOR) one~\cite{gor}, in the limit of
exact $SU(2)$ isospin symmetry it takes the form,
\be
\label{1}
m_{\pi}^2=-\frac{2m_q\langle\bar{q}q\rangle}{f_{\pi}^2},
\ee
where
$m_q$ is the current quark mass, $f_{\pi}=92.4$~MeV~\cite{pdg}
represents the weak pion decay constant, and
$\langle\bar{q}q\rangle$ is the quark condensate, the latter two
quantities are phenomenological parameters. Relation~\eqref{1}
comes from the interpretation of pion as the pseudogoldstone boson
of spontaneously broken chiral invariance of QCD in the limit
$m_q=0$. Since the term $m_q\bar{q}q$ enters the QCD Lagrangian,
the form of relation~\eqref{1} looks quite natural. In addition,
the linear dependence $m_{\pi}^2(m_q)$ can be obtained
heuristically --- since the relativistic field equations include
the linear fermion masses and the square boson ones, the simplest
nontrivial relation $m_{\pi}^2(m_q)$ is just a linear function.
However, $m_q$ does not represent a renorminvariant quantity while
$m_{\pi}^2$ does (it is an observable), but one can build a
renorminvariant quantity $m_q\langle\bar{q}q\rangle$ if the CSB
takes place ({\it i.e.} if $\langle\bar{q}q\rangle\neq0$), so
$m_{\pi}^2\sim m_q\langle\bar{q}q\rangle$. The relation
$m_{\pi}\sim f_{\pi}$ cannot take place as $f_{\pi}\neq0$ in the
limit $m_{\pi}=0$, $m_q=0$. Thus, one arrives at
relation~\eqref{1} just by dimensional analysis plus the
requirement of renorminvariance (factor 2 appears from the sum
$m_u+m_d$, the result can depend only on this sum in the linear
approximation by symmetry reasons). The linear dependence
$m_{\pi}^2=\Lambda m_q$ was checked on lattice~\cite{gorlat} and
up to all available $m_q$ in the lattice simulations the agreement
was quite prominent.

As to the spectrum of other light nonstrange mesons, the matter
did not go here beyond different models. In the first
approximation, the spectrum of masses square is equidistant,
\be
\label{2}
m_n^2=an+m_0^2, \qquad n=0,1,2,\dots,
\ee
where $n$
denotes either spin (the Regge trajectories) or "radial" quantum
number ({\it i.e.} it enumerates the daughter Regge trajectories),
the slope $a$ is nearly the same for both cases while the
intercept $m_0^2$ is channel-dependent. This picture agrees very
well with the phenomenology~\cite{ani} (see also discussions
in~\cite{sh,we,epj,mpla,bicudo}). The observations of
behaviour~\eqref{2} gave rise to the idea of strings and up to now
the conjecture that QCD in the hadronization regime is dual to
some string theory is still alive. In this case, the slope $a$ is
related to the string tension $\sigma$ as
\be
\label{3}
a=2\pi\sigma.
\ee
We note that the intercept $m_0^2$ should
contain a contribution arising from the term
$m_q\langle\bar{q}q\rangle$ since the arguments for its existence
are quite general. This is our first important observation.

We consider the spectrum of mesons as various excitations of the
pion. Since the (charged) pion lives near $10^{16}$ times as long
as its excitations, these excitations can be regarded as sudden
perturbations of the lightest bound state in QCD caused, say, by a
collision with a hadron. This hierarchy of time scales is our
second important observation: As is known from the Quantum
Mechanics the wave function of the bound state is not changed for
such a short period of time, hence, the relations involving the
bound state do not change.

\section{Model 1: Kaluza-Klein scenario}

Let us introduce one extra spatial dimension with the topology $S^1$
and radius $R$. According to the Kaluza-Klein theories (for a review see,
{\it e.g.},~\cite{kub}) a particle of mass $m$ excited along the 5-th
compactified dimension looks like a usual 4-dimensional particle with
the mass
\be
\label{4}
m_n^2=m^2+\frac{n^2}{R^2}, \qquad n=0,\pm1,\pm2,\dots
\ee

Assume now that one of quarks forming the lightest meson --- the
pion --- can be excited along the 5-th dimension and the
corresponding Kaluza-Klein modes give rise to the observed
spectrum of light nonstrange mesons (since we are dealing
semiclassically with a two-body system the excitation of both
quarks can be always represented as the excitation of one quark
only). Due to the linear dependence on quark mass,
$m_{\pi}^2=\Lambda m_q$ and the hierarchy of time scales
discussed in the previous Section,
relations~\eqref{1} and~\eqref{4} should
then dictate the following spectrum of excitations,
\be
\label{5}
m_n^2=-\frac{\langle\bar{q}q\rangle}{f_{\pi}^2}
\left(m_q+\sqrt{m_q^2+\frac{n^2}{R^2}}\right)+\tilde{m}_0^2,
\ee
where $\tilde{m}_0^2$ is a channel-dependent constant
parametrizing various side effects which could exert influence on
the determination of resonance position. In the light flavor
sector one can neglect the contribution of current quark mass
$m_q$ if $n\neq0$, the spectrum is then linear. Comparing the
slopes in Eqs.~\eqref{2} and~\eqref{5} we arrive at the following
matching condition,
\be
\label{6}
R=-\frac{\langle\bar{q}q\rangle}{f_{\pi}^2a}.
\ee
Thus, the size
of extra dimension and parameters of non-perturbative QCD turn out
to be remarkably related. For typical phenomenological values of
the quark condensate\footnote{The quantities $\langle\bar{q}q\rangle$
and $R$ are not renorminvariant, according to Eq.~\eqref{5}
renorminvariant is the combination $\langle\bar{q}q\rangle R^{-1}$. We
make estimates at the scale about 1~GeV and neglect the running of those
quantities with energy in the resonance region in what follows.}
and the slope,
$\langle\bar{q}q\rangle=-(0.235~\text{GeV})^3$ and
$a=(1.1~\text{GeV})^2$, relation~\eqref{6} yields the estimate
$R\approx1.3$~GeV$^{-1}$ (or 0.25~fm). This estimate and the fact that
$R$ is related to the order parameters of CSB suggest that
the compactification radius can be related to
$\Lambda_{\text{CSB}}$, $R\simeq\Lambda_{\text{CSB}}^{-1}$.

The validity of the idea above is supported by the following
estimate. The lifetime (inverse full decay width) of excited
mesons is just the time of "living" in the extra dimension. The
typical decay widths of hadron resonances show that their lifetime
is of the order of $\tau\sim10^{-24}$~s. Since the quarks inside
the light mesons are ultrarelativistic they propagate with the
speed of the order of $c=3\cdot10^8$~m/s. Hence, the size of extra
dimension can be estimated as $c\tau\sim0.3$~fm that is in accord
with our previous estimate.

\section{Model 2: Gluon string scenario}

Perhaps the most natural model of light mesons reproducing the
linear spectrum~\eqref{2} is the model of effective gluon string
--- a flux tube of chromoelectric field of constant density
$\sigma$ --- stretched between the quark an antiquark. Within such
models, spectrum~\eqref{2} is obtained as a result of
semiclassical quantization of the string (see, {\it
e.g.},~\cite{sh} and references therein). A drawback of these
models is that the size of the mesons grows linearly with the
mass, making the linear size of the highly excited states
unrealistically large while in reality this size is approximately
the same as the size of pions. To overcome the difficulty one may
assume that the gluon string is stretched not in the observable 3
spatial dimensions but in the extra dimension. If the string is
$n$ times wound around the compactified extra dimension, the
quark-antiquark system acquires the mass\footnote{In order to
avoid the chirality problem we must replace the circle formed by
the 5-th dimension by the orbifold --- a circle with identified
opposite points~\cite{kub}, in this way the physical interval
extends a length $\pi R$.} $M=\pi Rn\sigma=aRn/2$ (see
Eq.~\eqref{3}) which can be interpreted as an effective excitation energy
of one of quarks. Applying the arguments of the
previous Section we obtain that when the CSB occurs the spectrum
should be
\be
\label{02}
m_n^2=-\frac{\langle\bar{q}q\rangle}{f_{\pi}^2}
\left(2m_q+\frac{aRn}{2}\right)+\tilde{m}_0^2.
\ee
Matching of Eq.~\eqref{02} to Eq.~\eqref{2} yields
\be
\label{03}
R^{-1}\simeq-\frac{\langle\bar{q}q\rangle}{2f_{\pi}^2}\equiv \frac{B}{2},
\ee
where $B=-\frac{\langle\bar{q}q\rangle}{f_{\pi}^2}\approx1.5$~GeV
represents an important phenomenological parameter in the
low-energy effective theories of QCD.

Relation~\eqref{03} leads to the estimate $R\approx1.3$~GeV$^{-1}$.
Thus, the radius of compactified dimension in Model~2 is
practically equal to that of Model~1. This coincidence seems to be
not fortuitous --- it strongly suggests that both models are
complementary, hence, they can be matched to each other. Comparing
Eq.~\eqref{02} with Eq.~\eqref{5} and neglecting the current quark
mass $m_q$ we arrive at the relation
\be
\label{04}
a\simeq 2R^{-2}.
\ee
Substituting Eq.~\eqref{03} into Eq.~\eqref{04} we obtain
\be
\label{05}
\sqrt{a}\simeq-\frac{\langle\bar{q}q\rangle}{\sqrt{2}f_{\pi}^2}.
\ee
Relation~\eqref{05} shows explicitly how the spectrum of highly
excited states is determined by the order parameters of the chiral
symmetry breaking in QCD. It must be emphasized that the radius $R$
of compactified dimension disappears in the final expression~\eqref{05}.
This example demonstrates qualitatively how the extra dimensions can be
exploited as an auxiliary tool for finding new relations between
parameters of low-energy QCD.

\section{Alternative interpretations for compactification radius}

The interpretation of $R^{-1}$ as $\Lambda_{\text{CSB}}$ is
suggestive but not unique. Since according to our estimates
$R^{-1}\approx0.76$~GeV, it might be possible to identify $R^{-1}$
with the mass of $\rho$-meson,
$m_{\rho}\approx0.78$~GeV~\cite{pdg}.

Another tempting albeit highly speculative interpretation of $R$
in Model~1 consists in identifying $R^{-1}$ with the constituent
mass $M_{\text{con}}$. This interpretation could be deduced as
follows.

Instead of Eq.~\eqref{2} we should write
\be
\label{06}
m_n^2=m_{\rho}^2n,\qquad n=0,1,2,\dots,
\ee
where
the states alternate in parity. This is nothing but the spectrum
of the Ademollo--Veneziano--Weinberg dual amplitude~\cite{avw} and
is resembling the asymptotics of dim2 QCD spectrum in the
large-$N_c$ limit~\cite{dim2}. Simultaneously, one achieves
agreement in the slope value with the heavy-light quarkonia~\cite{ol}
as long as $m_{\rho}^2\simeq a/2$ numerically and in many models
(see, {\it e.g.},~\cite{plb}). Within Model~1,
the radius $R$ in Eq.~\eqref{6} is then enhanced by a factor of 2. We
observe then that the mass of the first Kaluza-Klein mode of light
quark (let be $m_q^{(1)}$) becomes close to the
constituent quark mass $M_{\text{con}}\simeq0.3$~GeV (say, for
$\langle\bar{q}q\rangle=-(0.25~\text{GeV})^3$ one has
$R\approx3$~GeV$^{-1}$ that gives $m_q^{(1)}\approx0.33$~GeV). The
identification of $m_q^{(1)}$ with $M_{\text{con}}$ provides a
rather unexpected prospects for the explanation of still enigmatic
success of nonrelativistic constituent quark model. In addition,
it suggests that since the $\rho$-meson is the first non-goldstone
excitation the formula for the $\rho$-meson mass is just the
GOR relation~\eqref{1} in which one of quark masses $m_q$ is
replaced by $M_{\text{con}}$,
\be
\label{10}
m_{\rho}^2\simeq-\frac{(m_q+M_{\text{con}})\langle\bar{q}q\rangle}{f_{\pi}^2}.
\ee
This relation is fulfilled with about 5-10\% accuracy,
in this regard it is interesting whether it could be
derived from some dynamical model. Relation~\eqref{10} is another
example of new relations which might by found with the help
of extra dimensions.


\section{Discussions}

Let us point out some important distinctions between the presented
scheme that we will refer to as Kaluza-Klein/QCD (KK/QCD) approach
(Model~1) and the AdS/QCD one. First of all, the AdS/QCD ideology
is motivated by the AdS/CFT correspondence~\cite{maldacena} while
we do not see such an inspiring analogue for the KK/QCD approach.
However, QCD is neither conformal no supersymmetric and for this
reason presently the AdS/QCD has a status of speculative
hypothesis that for some unknown reasons works surprisingly well
in description of hadron physics up to 2~GeV. In this regard, the
level of speculations in AdS/QCD and in KK/QCD is close. Second,
the AdS/QCD approach relates the QCD dynamics with metrics on the
boundary of AdS space that is induced by the gravity in the bulk.
In a sense, the KK/QCD approach is also related with a higher
dimensional gravity. The matter is that the KK-reduction of
gravity to four dimensions leads to appearance of radion (modulus
field) whose v.e.v. fixes the radius of extra
dimension~\cite{kub}. As within KK/QCD this radius is fixed by
relation~\eqref{6} we arrive at intriguing prospects to relate the
non-perturbative dynamics of strong QCD to various mechanisms of
modulus stabilization proposed in the literature (see, {\it
e.g.},~\cite{kub}). In a sense, the relation between AdS/QCD and
KK/QCD resembles the relation between the Rundall-Sundrum
models~\cite{rs} and the ADD models~\cite{add} of extra
dimensions. In both cases new high precision experiments in
spectroscopy of light hadrons below 2.5~GeV are needed to learn
about the underlying geometry.

Since the curvature of 5d space in KK/QCD is zero (or least is
negligible), there is an inspiring possibility that the extra
dimension represents a physical reality. If this is the case, the
KK/QCD scenario differs substantially from the ADD one. First, the
size of extra dimension is not sub-millimetric (up to 0.1~mm
according to the ADD-phenomenology), it is near 1~fm. This spoils
the main model-independent feature of the ADD-models
--- the prediction of high-multiplicity emission of light
Kaluza-Klein gravitons. Second, the excited quarks do not escape
into extra dimensions because their wavelength
$\lambda_n\sim1/m_n\sim R$ and the size of the compact dimension
are comparable, {\it i.e.} they represent genuine Kaluza-Klein
excitations with momentum in compact dimension, such states are
interpreted from the 4-dimensional point of view as particles of
mass $\sim 1/R$ that are still localized in our 4 dimensions. This
solves a possible problem with the conservation of electric charge
in our 4-dimensional world.

We have restricted ourselves by the mesons composed of light
nonstrange quarks. Concerning the other hadrons, the discussion
will be practically unchanged if we include light mesons
containing the strange quark since the GOR relation~\eqref{1}
remains valid and the corresponding Regge trajectories are also
linear with practically equal slope. As to the light baryons,
their spectrum can be well fitted by a formula like~\eqref{2} with
the same slope~\cite{kl}. This gives a certain hope that the dual
Kaluza-Klein mechanism is also relevant here, but the situation
may be more tricky since we do not have any GOR-like relation for
the ground state mass. In the heavy-light quarkonia, one could
expect similar Kaluza-Klein excitations of the light constituent.
However, the slope of the corresponding trajectories seems to be
half of the value for the light-light quarkonia~\cite{ol} (see,
however, the discussions of Eq.~\eqref{06}). Concerning the rest
of the hadrons, we have nothing to say presently.

\section{Conclusions}

On a rather heuristic level, we have discussed possible
geometrical interpretations of such non-perturbative aspects of
QCD as the chiral symmetry breaking and the meson spectrum. It is
argued that the effective theories for strong QCD could be
searched on the base of 5d field theories with one compactified
extra dimension. In the region of low and intermediate energies,
such models represent an alternative to the usual AdS/QCD
bottom-up approach where the extra dimension is warped. Within the
conjectured scenario, the abundance of hadron resonances composed
of light quarks is interpreted as the effect of the Kaluza-Klein
excitations of the quarks inside mesons or, alternatively, as the
formation of gluon strings wound around compactified extra
dimension. The ensuing models entail some simple relations between
the size of extra dimensions and phenomenological quantities
characterizing the non-perturbative QCD. The matching of this
picture to the phenomenology is achieved for a typical size of
extra dimension of the order of 1~fm. If the 5th dimension
represents a physical reality the outlined picture suggests that
its size and possibly another characteristics could be probed by
high precision experiments in hadron spectroscopy below 2.5~GeV.

Unfortunately, the presented arguments are not yet backed up by a
careful theoretical analysis of a specific field theory. It would
be extremely interesting to construct an explicit example of such
a theory.

\section*{Acknowledgments}

The work is supported by the Alexander von Humboldt Foundation and
by RFBR, grant no. 09-02-00073-a.


\begin{thebibliography}{99}
\bibitem{maldacena} J. M. Maldacena, Adv. Theor. Math. Phys. {\bf 2}, 231 (1998).
\bibitem{hw} J.~Polchinski and M.~J.~Strassler, Phys. Rev. Lett. {\bf 88}, 031601
(2002); H.~Boschi-Filho and N.~R.~F.~Braga, Eur. Phys. J. C {\bf
32}, 529 (2004); JHEP {\bf 0305}, 009 (2003); G.~F.~de Teramond
and S.~J.~Brodsky, Phys. Rev. Lett.  {\bf 94}, 201601 (2005);
J.~Erlich, E.~Katz, D.~T.~Son and M.~A.~Stephanov, Phys. Rev.
Lett. {\bf 95}, 261602 (2005); L.~Da Rold and A.~Pomarol, Nucl.
Phys. B {\bf 721}, 79 (2005).
\bibitem{AdS} A. Karch, E. Katz, D. T. Son, and M. A. Stephanov,
Phys. Rev. D {\bf 74}, 015005 (2006);  H. Forkel, M. Beyer, and T.
Frederico,  JHEP {\bf 0707}, 077 (2007).
\bibitem{erd} J. Erdmenger, N. Evans, I. Kirsch, and E. Threlfall,
Eur. Phys. J. A {\bf 35}, 81 (2008).
\bibitem{er} J. Erlich and C. Westenberger, arXiv:0812.5105 [hep-ph];
T.~D.~Cohen, arXiv:0805.4813 [hep-ph].
\bibitem{kub} R. Sundrum, hep-th/0508134.
\bibitem{afonin} S.~S.~Afonin, arXiv:0902.3959 [hep-ph].
\bibitem{gor} M. Gell-Mann, R. J. Oakes, and B. Renner, Phys. Rev. {\bf 175}, 2195 (1968).
\bibitem{pdg} Particle Date Group (C. Amsler {\it et al.}), Phys. Lett. B {\bf 667}, 1 (2008).
\bibitem{gorlat} S. J. Dong, F. X. Lee, K. F. Liu, and J. B.
Zhang, Phys. Rev. Lett. {\bf 85}, 5051 (2000).
\bibitem{ani} A. V. Anisovich, V. V. Anisovich, and A. V. Sarantsev,
Phys. Rev. D~{\bf 62}, 051502(R) (2000);
V.~V.~Anisovich, Phys. Usp. {\bf 47}, 45 (2004);
D.~V.~Bugg, Phys. Rept. {\bf 397}, 257 (2004).
\bibitem{sh} M. Shifman and A. Vainshtein, Phys. Rev. D~{\bf 77}, 034002 (2008).
\bibitem{we} S. S.~Afonin, A. A.~Andrianov, V. A.~Andrianov, and D.~Espriu,
JHEP {\bf 0404}, 039 (2004).
\bibitem{epj} S.~S. Afonin, Eur. Phys. J. A {\bf 29}, 327 (2006);
Phys. Lett. B {\bf 639}, 258 (2006).
\bibitem{mpla} S.~S. Afonin, Phys. Rev. C {\bf 76}, 015202 (2007);
Mod. Phys. Lett. A {\bf 22}, 1359 (2007);
Int. J. Mod. Phys. A {\bf 22}, 4537 (2007).
\bibitem{bicudo} P. Bicudo, Phys. Rev. D {\bf 76}, 094005 (2007);
S.~S.~Afonin, Int. J. Mod. Phys. A {\bf 23}, 4205 (2008).
\bibitem{avw} M. Ademollo, G. Veneziano, and S. Weinberg,
Phys. Rev. Lett. {\bf 22}, 83 (1969).
\bibitem{dim2} G.~'t Hooft, Nucl. Phys. B {\bf 75}, 461 (1974);
C. G. Callan, N. Coote, and D. J. Gross, Phys. Rev. D~{\bf 13},
1649 (1976); M. B. Einhorn, Phys. Rev. D~{\bf 14}, 3451 (1976).
\bibitem{ol} M. G. Olsson, Phys. Rev. D~{\bf 5}, 5479 (1997);
Yu. S. Kalashnikova, A. V. Nefediev, and Yu. A. Simonov,
Phys. Rev. D~{\bf 64}, 014037 (2001).
\bibitem{plb} S. S. Afonin, Phys. Lett. B {\bf 576}, 122 (2003);
Nucl. Phys. B~{\bf 779}, 13 (2007);
S. S. Afonin and D. Espriu, JHEP {\bf 0609}, 047 (2006).
\bibitem{rs} L. Randall and R. Sundrum, Phys. Rev. Lett.~{\bf 83}, 3370
(1999); {\it ibid.}~{\bf 83}, 4690 (1999).
\bibitem{add} N. Arkani-Hamed, S. Dimopoulos, and G. Dvali, Phys. Lett. B {\bf 429}, 263
(1998); Phys. Rev. D {\bf 59}, 086004 (1999).
\bibitem{kl} E. Klempt, Phys. Rev. C {\bf 66}, 058201 (2002).
\end{thebibliography}
\end{document}